%% file: main_v2.tex
\documentclass[aps,prx,superscriptaddress,twocolumn]{revtex4-2}
\usepackage[utf8]{inputenc}
\usepackage{bm}
\usepackage{bbm}
\usepackage{bbold}
\usepackage{amsfonts}
\usepackage{amsmath}
\usepackage{amssymb}
\usepackage{graphicx}
\usepackage{mathtools}
\usepackage{upgreek}
\usepackage{xfrac}
\usepackage{soul}
\usepackage{xcolor}

\usepackage[colorlinks=true,citecolor=blue]{hyperref}
\hypersetup{colorlinks=true,citecolor=blue,linkcolor=blue,urlcolor=blue}
\usepackage{url}

\usepackage{txfonts}
\DeclareSymbolFont{matha}{OML}{txmi}{m}{it}% txfonts
\DeclareMathSymbol{\varv}{\mathord}{matha}{118}

\makeatletter
\newsavebox{\@brx}
\newcommand{\llangle}[1][]{\savebox{\@brx}{\(\m@th{#1\langle}\)}%
  \mathopen{\copy\@brx\kern-0.5\wd\@brx\usebox{\@brx}}}
\newcommand{\rrangle}[1][]{\savebox{\@brx}{\(\m@th{#1\rangle}\)}%
  \mathclose{\copy\@brx\kern-0.5\wd\@brx\usebox{\@brx}}}
\makeatother

\begin{document}

\title{What can we learn from quantum convolutional neural networks?}

\author{Chukwudubem Umeano}
\affiliation{Department of Physics and Astronomy, University of Exeter, Stocker Road, Exeter EX4 4QL, United Kingdom}

\author{Annie E. Paine}
\affiliation{Department of Physics and Astronomy, University of Exeter, Stocker Road, Exeter EX4 4QL, United Kingdom}
\affiliation{PASQAL, 7 Rue Léonard de Vinci, 91300 Massy, France}

\author{Vincent E. Elfving}
\affiliation{PASQAL, 7 Rue Léonard de Vinci, 91300 Massy, France}

\author{Oleksandr Kyriienko}
\affiliation{Department of Physics and Astronomy, University of Exeter, Stocker Road, Exeter EX4 4QL, United Kingdom}
\affiliation{PASQAL, 7 Rue Léonard de Vinci, 91300 Massy, France}

\date{\today}

\begin{abstract}
Quantum machine learning (QML) shows promise for analyzing quantum data. A notable example is the use of quantum convolutional neural networks (QCNNs), implemented as specific types of quantum circuits, to recognize phases of matter. In this approach, ground states of many-body Hamiltonians are prepared to form a quantum dataset and classified in a supervised manner using only a few labeled examples. However, this type of dataset and model differs fundamentally from typical QML paradigms based on feature maps and parameterized circuits. In this study, we demonstrate how models utilizing quantum data can be interpreted through hidden feature maps, where physical features are implicitly embedded via ground-state feature maps. By analyzing selected examples previously explored with QCNNs, we show that high performance in quantum phase recognition comes from generating a highly effective basis set with sharp features at critical points. The learning process adapts the measurement to create sharp decision boundaries. Our analysis highlights improved generalization when working with quantum data, particularly in the limited-shots regime. Furthermore, translating these insights into the domain of quantum scientific machine learning, we demonstrate that ground-state feature maps can be applied to fluid dynamics problems, expressing shock wave solutions with good generalization and proven trainability.
\end{abstract}

\maketitle

\section{Introduction}

Quantum computing offers a paradigm for solving computational problems in a distinct way \cite{nielsen2010quantum,Montanaro2016}. It has been considered for addressing challenges in chemistry \cite{McArdle2020RMP,elfving2020quantum}, material science \cite{Barends2015,Ebadi2021,Bespalova2021,Stanisic2022}, high-energy physics \cite{Banuls2020,Haase2021resourceefficient}, optimization and finances \cite{Egger2020,leclerc2022financial,Herman2023}, and recently, for solving machine learning problems on quantum computers \cite{Schuld2015rev,Biamonte2017,Benedetti2019rev,PerdomoOrtiz2018rev,Dunjko2020nonreviewofquantum}. The latter is the task of quantum machine learning (QML). Quantum machine learning is a rapidly progressing field of research, which comprises different techniques that may offer a speedup \cite{Biamonte2017}, as well as a range of other advantages that have not been considered before \cite{Schuld2022PRXQ,Dunjko2018rev}. To date, this includes examples from supervised learning (represented by classification \cite{Rebentrost2014PRL,farhi2018classification,Grant2018,Havlicek2019,henderson2019quanvolutional,Schuld2020PRA,chen2021hybrid,Belis2021,Nghiem2021,SChen2022,Schetakis2022,Jaderberg2022} and regression \cite{Wiebe2012,Schuld2016PRA,GWang2017,mitarai2018quantum,kyriienko2021solving,paine2023quantum}), reinforcement learning \cite{Dunjko2016PRL,Melnikov2018,SChen2020,Saggio2021}, and unsupervised learning with a strong effort in generative modelling \cite{Benedetti2019rev,Liu2018,PerdomoOrtiz2018,Coyle2020,Zoufal2019,Paine2021,kyriienko2022protocols,GiliPRA2023,gili2023generative}. By far the strongest effort concerns classification \cite{Li2021rev}. A typical workflow starts with loading classical datasets $\mathcal{D} = \{ \bm{x}_m, y_m\}_{m=1}^{M}$ comprising $M$ data samples. The features $\bm{x}$ can be embedded into parametrized gates or quantum state amplitudes, and by processing the corresponding quantum states one can match class labels $y_m$ from the training subset \cite{Benedetti2019rev}. The goal is to predict unseen samples. Here, the workflow involves quantum circuits for \emph{embedding} classical data (known as quantum feature maps $\hat{\mathcal{U}}_{\varphi}(\bm{x})$ \cite{Schuld2019feature,mitarai2018quantum,SchuldSweke2021PRA}), which generate a state in Hilbert space of the processing device, $\bm{x} \mapsto |\psi(\bm{x})\rangle = \hat{\mathcal{U}}_{\varphi}(\bm{x})|{\o}\rangle$. The states are then processed by variational circuits (aka ansatze) $\hat{\mathcal{V}}_{\bm{\theta}}$ \cite{Benedetti2019rev,Cerezo2021rev}, and measured for some observable $\hat{\mathcal{O}}$. This distantly resembles deep learning \cite{LeCun2015}, for the model formed in a quantum latent space, and is referred to as a quantum neural network (QNN) \cite{SchuldSweke2021PRA,Abbas2021,Goto2021PRL}. While QNNs were successfully applied to many proof-of-principal tasks \cite{Cerezo2021rev,Li2021rev}, some issues remain to be resolved before seeing their utility in practice. These issues include limited trainability for models of increased size, corresponding to barren plateaus (BPs) of vanishing gradients of QNNs \cite{mcclean2018barren,Cerezo2021NatComm}, and generally rugged optimization landscape \cite{Anschuetz2022,Wang2022energylandscapeof}. Also, despite an increased expressive power of QNN-based models \cite{Du2020PRRes,Abbas2021,Du2021PRXQ}, they remain heuristic in nature, and a clear separation of quantum and classical model performance can only be established for very peculiar datasets \cite{Liu2021NatPhys,Huang2021NatComm}. Another strongly related issue is that high expressivity of quantum models implies limited trainability \cite{Holmes2022PRXQ}. This may lead to overfitting in cases where the chosen basis set does not match the required problem \cite{Caro2021encodingdependent}.%, and needs to balanced for making models efficiently trainable.
%%%
\begin{figure*}[t!]
\includegraphics[width=1.0\linewidth]{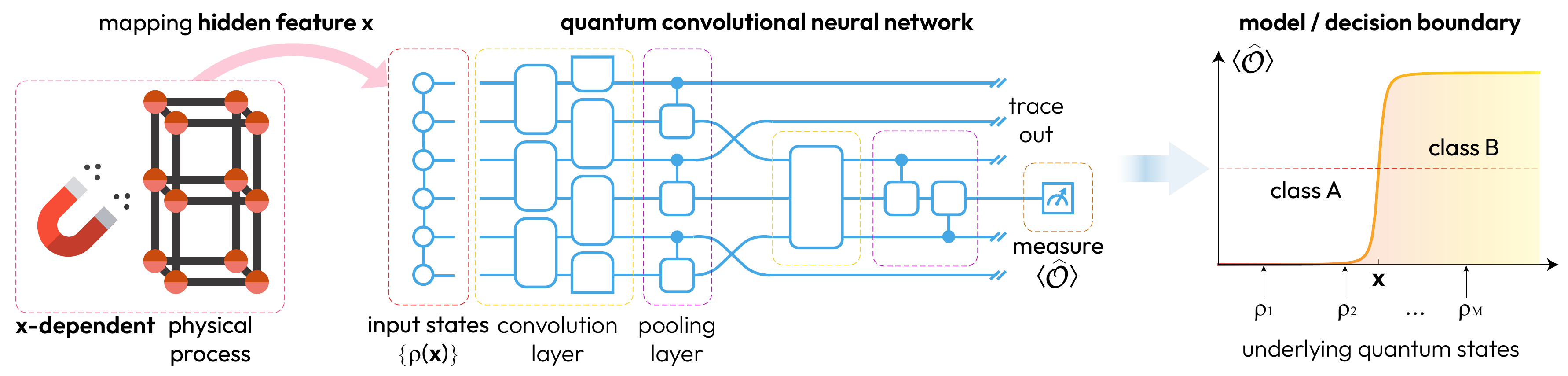}
    \caption{Visualization for the classification process of physical phases with a quantum convolutional neural network (QCNN). We highlight that input states $\hat{\rho}(\bm{x})$ for the network come from actual physical processes, e.g. preparation of ground states for spin lattices. The analyzed states depend on underlying classical features $\bm{x}$ of the system, being the physical parameters (externally controlled like magnetic field and temperature, or internal parameters like degree of anisotropy). This can be seen as a \emph{hidden} feature map (left). The goal of QCNN is then building a model based on a simple few-qubit observable $\hat{\mathcal{O}}$, with its expectation $\langle \hat{\mathcal{O}}\rangle (\bm{x})$ representing a nonlinear decision boundary with respect to the system features.}
    \label{fig:hidden-map}
\end{figure*}
%%%

Recently, an increasing number of works considered quantum machine learning models for training on \emph{quantum} data~\cite{Huang2022,Caro2022,schatzki2021entangled,Larocca2022PRXQ,nguyen2022theory}. In this case a quantum dataset $\mathcal{Q} = \{ \hat{\rho}_\alpha, y_\alpha \}_{\alpha=1}^{M}$ corresponds to the collection of $M$ quantum states $\hat{\rho}_\alpha$ (pure or mixed) coming from some quantum process, and associated labels $y_\alpha$ that mark their distinct class or property. When learning on quantum data, quantum computers have shown excellent results in sample complexity \cite{Caro2022,Huang2021PRL,Huang2022,huang2022foundations}. A striking example corresponds to quantum convolutional neural networks (QCNNs). Introduced in Ref.~\cite{Cong2019} by Cong \textit{et al.}, this type of network was designed to take quantum states, and using \emph{convolution} (translationally-invariant variational quantum circuits) and \emph{pooling} (measurement with conditional operations) prepare an efficient model for predicting the labels. This was used for quantum phase classification at increasing scale applied to spin-1/2 cluster model and Ising model in one-dimensional (1D) systems. Additionally, it was implemented in superconducting circuit hardware, classifying 1D cluster model states. The results for phase classification were reproduced and extended in other studies \cite{Caro2022,LiuPollmann2023qcnn,goh2023liealgebraic,herrmann2022realizing,zapletal2023error} showing that few samples are needed for training, generalization is excellent \cite{Caro2022}, and accuracy is superior to other approaches \cite{goh2023liealgebraic}. Moreover, the scaling of gradients was shown to decrease polynomially in the system size (thus --- efficiently), such that QCNNs avoid barren plateaus \cite{pesah2021absence}. This is in contrast to exponentially decreasing gradient variance for generic deep circuits \cite{mcclean2018barren,Cerezo2021NatComm}. Current intuition is that working with quantum data leads to better generalization, as QCNNs inherently ``analyze'' the structure of entanglement of quantum states \cite{Cong2019}, and reduced entanglement leads to better gradients \cite{Patti2021,Sack2022PRXQ}. However, despite the success of QCNNs, their full understanding is missing, and leaves them aside of mainstream QML. In particular, the source of the good generalization for these models remains unclear \cite{gil2024understanding}. This hinders the development and use of QCNN-type models beyond explored classification examples.

In this work, we aim to demystify the inner workings of quantum convolutional neural networks, specifically highlighting why they are so successful in quantum phase recognition. We show that supplied quantum states (features) can be understood in terms of \emph{hidden} feature maps --- quantum processes that prepare states depending on classical parameters $\bm{x}$ (being a feature vector or scalar; see Fig.~\ref{fig:hidden-map}). During the mapping process, we observe that the ground state preparation supplies a very beneficial basis set, which allows the building of a nonlinear quantum model for the decision boundary --- the ultimate goal of classification --- that is sharp and generalizes on few data points and with smaller number of measurement shots. This is compared to the rotation-based Fourier embedding, which requires feature engineering for sufficient generalization. We note that the utility of ground state embeddings is not limited to QCNNs, but could be an avenue to improved performance for general quantum models. We connect the developed QCNN description with error correction and multi-scale entanglement renormalization ansatz (MERA) explanations from Ref.~\cite{Cong2019}, and show that single-qubit observables can be used to ``pick up'' the most suitable basis functions, while leading to sampling advantage. Motivated by classification, we apply the described QCNN workflow with ground state embedding to solve regression problems motivated by fluid dynamics. This can offer QML tools to deal with critical phenomena with improved generalization. 

%=====================

\section{Background}

\subsection{Quantum neural networks as Fourier-type models}

Quantum machine learning has evolved from being perceived as a linear algebra accelerator \cite{Biamonte2017} into a versatile tool for building quantum models. The typical workflow corresponds to mapping (embedding) a classical dataset $\mathcal{D}$ with a feature map as a quantum circuit $\hat{\mathcal{U}}_{\varphi}(x)$ such that features $x \mapsto |\psi(x)\rangle = \hat{\mathcal{U}}_{\varphi}(x) |{\o}\rangle$ are represented in the latent space of quantum states \cite{SchuldSweke2021PRA,Goto2021PRL,kyriienko2021solving}. Here $|{\o}\rangle \coloneqq |0\rangle^{\otimes N}$ is the computational zero state. The next step corresponds to adapting the generated states with a variational circuit $\hat{\mathcal{V}}_{\theta}$, and reading out the model as an expectation value of an observable $\hat{\mathcal{O}}$. The last two steps can also be seen as a measurement adaptation process \cite{schuld2021supervised}. Summarizing the workflow, we build QML models as 
\begin{equation}
\label{eq:QNN-model}
    f_{\theta}(x) = \langle {\o}| \hat{\mathcal{U}}_{\varphi}(x)^\dagger \hat{\mathcal{V}}_{\theta}^\dagger  \hat{\mathcal{O}}  \hat{\mathcal{V}}_{\theta} \hat{\mathcal{U}}_{\varphi}(x) |{\o} \rangle,
\end{equation}
 The corresponding approach for instantiating $f_{\theta}(x)$ is often referred to as the quantum neural network approach. Usually, the embedding is performed by circuits that involve single-qubit rotations $\hat{R}_{\beta} = \exp[-i x \hat{P}_{\beta}/2]$ \cite{mitarai2018quantum,Schuld2019parametershift} (for Pauli matrices $\hat{P}_{\beta} = \hat{X}, \hat{Y}, \hat{Z}$) or evolution of some multi-qubit Hamiltonian $\hat{G}$ (being the generator of dynamics) such that the map is $\hat{\mathcal{U}}_{\varphi}(x) = \exp[-i x \hat{G}/2]$. Following this convention, it is easy to see that under the spectral decomposition of $\hat{G} = \sum_j \lambda_j |s_j\rangle\langle s_j|$ (with eigenvalues $\{ \lambda_j \}_j$ and eigenstates $\{ |s_j\rangle \}_j$) the feature map includes complex exponents that depend on $x$ \cite{SchuldSweke2021PRA,kyriienko2021generalized}. Importantly, for the specified structure of the feature map the transformation to the diagonal basis is $x$-independent. When accounting for the structure of expectation values, the differences $\{ |\lambda_j - \lambda_{j'}| \}_{j,j'}$ of eigenvalues (spectral gaps) appear as frequencies for the underlying Fourier basis. The action of the variational circuit is then to ``weight'' the Fourier components, but not the features directly. The resulting model can be written as~\cite{SchuldSweke2021PRA}
\begin{equation}
\label{eq:Fourier-model}
    f_{\theta}(x) = \sum\limits_{\omega \in \Omega} c_{\omega}(\theta) e^{i \omega x}, 
\end{equation}
where $c_{\omega}(\theta)$ are coefficients that depend on matrix elements of $\hat{\mathcal{O}}$ and $\hat{\mathcal{V}}_{\theta}$, and $\Omega$ denotes a finite bandwidth spectrum of frequencies generated by the feature map. Its degree depends on the generator $\hat{G}$ and its eigenvalues. In Ref.~\cite{SchuldSweke2021PRA} the authors mention that rescaling of $x$ as $\varphi(x)$ does not change the model qualitatively, and we can see QNNs as Fourier-type models of large size \cite{schuld2021supervised}. Within this picture one can even imagine randomized Fourier models that can be treated classically and have similar performance \cite{landman2022classically}. 

Looking back into the steps leading to Eq.~\eqref{eq:Fourier-model}, we highlight that the presented description is by no means a complete guide to QNNs and building of quantum machine learning model in general. We observe that it implies the crucial assumption of the unitarity of the feature map and its form $\exp(-ix\hat{G}/2)$ that generates exponents as basis functions. Recently, the embeddings based on linear combinations of unitaries (LCU) were proposed that break this assumption \cite{williams2023quantum}. This is represented by the orthogonal Chebyshev feature map $\hat{\mathcal{U}}_{\varphi}(x)$ such that it generates states of the form $|\tau(x)\rangle = \sum_k c_k T_k(x)|k\rangle$, where $T_k(x)$ are amplitudes corresponding to Chebyshev polynomials of first-kind and degree $k$, and $c_k$ are constant factors. Importantly, here the amplitudes are $x$-dependent and form the basis for future quantum modelling. Another counter example to the Fourier-type models is the embedding of the type
\begin{equation}
\label{eq:noncommuting-map}
    \hat{\mathcal{U}}_{\varphi}(x) = \exp\left (-\frac{i}{2} \hat{G}_0 - \frac{i x}{2} \hat{G}_1 \right),
\end{equation}
where $[\hat{G}_0, \hat{G}_1] \neq 0$ as generators do not commute. In this case the spectral representation of the generator $\hat{G}(x) \coloneqq \hat{G}_0 + x \hat{G}_1 $ requires a basis transformation $\hat{W}(x)$ that is $x$-dependent. This leads to a feature dependence appearing in the coefficients $c_{\omega}(\theta,x)$ in Eq.~\eqref{eq:Fourier-model}, and departs from the Fourier modelling in a more nontrivial way than simply rescaling, $x \rightarrow \varphi(x)$. In cases of LCU and noncommuting embedding, or indeed any other case that does not fit the evolution embedding, the generated quantum states shall be seen as feature-dependent states
\begin{equation}
    |\psi(x)\rangle = \sum_j \phi_j(x)|j\rangle,
\end{equation}
with amplitudes $\{\phi_j(x)\}$ of orthonormal states $\{|j\rangle \}$ being functions of $x$, and representing the basis for building quantum models.

Given this layout of quantum neural networks, an interesting question arises: how does the Fourier model description fit the quantum data story and QCNN-based models that are too distinct from what we just described?

%------------

\subsection{Quantum convolutional neural networks: prior art}
In the seminal paper by Cong \textit{et al.} \cite{Cong2019} the authors have put forward the model with convolution and pooling layers, suggested as an analog of classical convolutional neural networks. They have used it for processing quantum states $|\psi_{\alpha}\rangle$ (or $\hat{\rho}_\alpha = |\psi_{\alpha}\rangle \langle \psi_{\alpha}|$) as ground states of spin models. The circuit consists of convolution unitaries $\{ \hat{U}_i \}$ and controlled unitaries $\{ \hat{V}_i \}$ for pooling (both considered being adjustable), followed by measuring $\hat{\mathcal{O}}$ as a few-qubit observable, while discarding the rest of the qubit register (collapsed on some measurement outcomes). The model then becomes
\begin{equation}
\label{eq:QCNN-model}
    f_{\{\hat{U}_i, \hat{V}_i,\hat{\mathcal{O}}\}}(|\psi_\alpha\rangle) = \langle \psi_\alpha| \prod\limits_{i=L}^{1}(\hat{U}_i^\dagger \hat{V}_i^\dagger) \hat{\mathcal{O}} \prod\limits_{i=1}^{L}( \hat{V}_i \hat{U}_i)|\psi_\alpha\rangle,
\end{equation}
where we again stress that $\hat{\mathcal{O}}$ corresponds to measuring $\tilde{N} \ll N$ qubits, while tracing out the rest. Here, unitaries $\{ \hat{U}_i, \hat{V}_i \}$ can be varied with $O(1)$ variational parameters, and layers are translationally invariant. Then the QCNN has only $O(\log N)$ variational parameters for $N$-qubit models, and is trainable \cite{pesah2021absence}. The task is to take the model in Eq.~\eqref{eq:QCNN-model} and fit it to label values of the quantum dataset $\{ y_\alpha\}$ for each state. This can be achieved by optimizing the mean squared error (MSE) loss
\begin{equation}
\label{eq:QCNN-MSE-loss}
    \mathcal{L}_{\mathrm{MSE}} = \frac{1}{M} \sum\limits_{\alpha = 1}^{M} \left\{ y_\alpha - f_{\{\hat{U}_i, \hat{V}_i,\hat{\mathcal{O}}\}}(|\psi_\alpha\rangle) \right\}^2.
\end{equation}
The loss can be minimized via the gradient descent or any other method suitable for non-convex optimization. As for the dataset, it was suggested to use ground states of spin-1/2 Hamiltonians. In particular, in most QCNN studies the cluster Hamiltonian with magnetic field and Ising terms was considered \cite{Cong2019,Caro2022}, corresponding to
\begin{equation}
\label{eq:H_QCNN}
    \hat{\mathcal{H}}_{\mathrm{QCNN}} = -J \sum\limits_{i=1}^{N-2} \hat{Z}_{i} \hat{X}_{i+1} \hat{Z}_{i+2} - h_{\mathrm{x}} \sum\limits_{i=1}^{N} \hat{X}_{i} - J_{\mathrm{xx}} \sum\limits_{i=1}^{N-1} \hat{X}_{i} \hat{X}_{i+1},
\end{equation}
where open boundary conditions are considered. This specific Hamiltonian was chosen as an example of non-trivial spin order in 1D, being uniquely related to measurement-based quantum computing \cite{raussendorf2022measurementbased}. In Eq.~\eqref{eq:H_QCNN} the first term corresponds to the spin-1/2 cluster Hamiltonian with three-body terms, the second term represents the transverse field, and the third term contains Ising interaction terms. The point $h_{\mathrm{x}} = J_{\mathrm{xx}} = 0$ corresponds to a $\mathbb{Z}_2 \times \mathbb{Z}_2$ symmetric Hamiltonian that hosts in its ground state a symmetry protected topological (SPT) phase. For $J_{\mathrm{xx}} = 0$ one can study the transition from SPT order at $J > h_{\mathrm{x}}$ to staggered ferromagnetic order at $J < h_{\mathrm{x}}$, with $h_{\mathrm{x}}/J = 1$ corresponding to the critical point. In the SPT phase and for the open boundary the ground state is four-fold degenerate, and is characterized by the string order parameter and nonzero topological entanglement entropy \cite{zeng2018quantum}. Examples of QCNN-based classification show that by measuring the string order $\hat{\mathcal{O}} = \hat{Z}_1 \hat{X}_2 \hat{X}_3 ... \hat{X}_{N-2 } \hat{X}_{N-1} \hat{Z}_N$ one can distinguish the SPT phase from other phases (open boundary). It is also known that string order implies the utility of states for performing measurement-based quantum computing \cite{raussendorf2022measurementbased} for which the ground state subspace (cluster states) represent the resource. This model is interesting to study due to the several transitions that can be identified. However, from the QML perspective the introduction of critical lines makes the analysis of basis functions opaque, so we simplify the model in our following study.

From the perspective of ground state analysis, the QCNN circuit is designed to measure the effective decision boundary as an expanded string order \cite{Cong2019}, and the pre-defined QCNN circuit can do this exactly if it satisfies certain criteria (see Ref.~\cite{Cong2019}, section ``Construction of QCNN'' in Methods), being suggested as guiding principles for building QCNNs. The first guiding principle is referred to as the fixed point criterion, where the exact cluster state $|\psi_0\rangle_N$ for $N$ qubits has to be convoluted and pooled to the $|\psi_0\rangle_{N/3}$ cluster state, with measurements of $2N/3$ qubits deterministically giving $0$ bits upon measurement. The second guiding principle is named as the error correction criterion, where pooling layers have to be designed such that errors that commute with global symmetry are fixed during measurements. This procedure is related to MERA. However, from the point of view of QML, it is yet to be understood how does entanglement play a role in this workflow.

%==========

\section{Model}

Here, we take the QCNN structure for quantum phase recognition and connect it with the standard QML workflow. For this we note that quantum states $|\psi_\alpha\rangle$ are not abstract quantum ``data points'', but in fact are the result of quantum processes that correspond to thermalization or ground state (GS) preparation. Specifically, they depend on properties of the underlying Hamiltonian, and its parameters representing features $x$. Namely, for the cluster-type Hamiltonian used in QCNNs so far these are couplings $J$, $J_{\mathrm{xx}}$ and magnetic field $h_{\mathrm{x}}$. These system parameters represent a feature vector $x = \{ J, h_{\mathrm{x}}, J_{\mathrm{xx}} \}$ (and each concrete realization is denoted by index $\alpha$). Even though we may not have access to them, they define the underlying behavior of the system we attempt to classify. The process of embedding these features as a part of the state preparation we call a \emph{hidden} feature map (Fig.~\ref{fig:hidden-map}). The circuits for preparing QCNN input as ground states of models $|\psi_0(x)\rangle$ may differ, and can come both from experiments (e.g. sensors) or specifically designed GS preparation schedule. In the following we assume that there exists a unitary $\hat{\mathcal{U}}_{\varphi}(x)$ or a completely positive trace-preserving map $\mathcal{E}_{\varphi}(x)$ for the ground state preparation of studied Hamiltonians, and summarize different options in the Appendix A.

%----

\subsection{Cluster ground state embedding}

Let us now consider an example that can shed light on the internal structure of quantum models working with quantum data. For this we take a spin-1/2 cluster Hamiltonian with periodic boundary. For understanding essential parts of the workflow, we prefer to set $J_\mathrm{xx}=0$ such that only one transition point and one variable function are analyzed. In this case the ground state in the SPT phase is unique and non-degenerate. From a technical perspective this enables working at smaller system size without being harmed by finite size effects, and avoid thermal ensemble state preparation (as pure ground state suffices in this case). Specifically, we consider the Hamiltonian of the form
\begin{equation}
\label{eq:H_t-cluster}
    \hat{\mathcal{H}}_{\mathrm{t-cluster}} = -J \sum\limits_{i=1}^{N} \hat{Z}_{i} \hat{X}_{i+1} \hat{Z}_{i+2} - h_{\mathrm{x}} \sum\limits_{i=1}^{N} \hat{X}_{i} - h_{\mathrm{z}} \sum\limits_{i=1}^{N} \hat{Z}_{i},
\end{equation}
where we have introduced an additional weak symmetry breaking term with a longitudinal field $h_{\mathrm{z}} \ll h_{\mathrm{x}},J$. We mostly care about the $J/h_{\mathrm{x}}$ transition, and for practical reasons keep $h_{\mathrm{z}} = 10^{-2} J$. This ensures that we break the degeneracy between states that have $\mathbb{Z}_2$ symmetry. Note that in the Hamiltonian $\hat{\mathcal{H}}_{\mathrm{t-cluster}}$ [Eq.~\eqref{eq:H_t-cluster}] we use periodic boundary conditions such that $\hat{X}_{N+1} \equiv \hat{X}_1$, $\hat{Z}_{N+1} \equiv \hat{Z}_1$, $\hat{Z}_{N+2} \equiv \hat{Z}_2$ etc. Finally, it is convenient to reparametrize the transverse cluster Hamiltonian in the form
\begin{equation}
\label{eq:H(x)}
    \hat{\mathcal{H}}(x) = -\cos\left( \frac{\pi x}{2}\right) \sum\limits_{i=1}^{N} \hat{Z}_{i} \hat{X}_{i+1} \hat{Z}_{i+2} - \sin\left( \frac{\pi x}{2}\right) \sum\limits_{i=1}^{N} \hat{X}_{i} - \varepsilon \sum\limits_{i=1}^{N} \hat{Z}_{i},
\end{equation}
where we introduced the effective parameters
\begin{equation}
\label{eq:x-def}
    x \coloneqq \frac{2}{\pi} \arcsin\left(\frac{h_{\mathrm{x}}}{\sqrt{J^2 + h_{\mathrm{x}}^2}}\right), ~~ \text{and} ~~ \varepsilon \coloneqq \frac{h_{\mathrm{z}}}{\sqrt{J^2 + h_{\mathrm{x}}^2}} \ll 1.
\end{equation}
We assume that $x$ changes from 0 to 1 via control of $h_x$, while $h_z$ is adjusted to keep $\varepsilon$ constant and small (ensuring a finite but small symmetry breaking term). Next, we prepare the ground state (GS) of $\hat{\mathcal{H}}(x)$ assuming one of the preparation strategies (Appendix A). The corresponding circuit, which we denote $\hat{\mathcal{U}}_{\varphi}(x)$, represents our feature map, acting such that $\hat{\mathcal{U}}_{\varphi}(x)|{\o}\rangle \eqqcolon |\psi_0(x)\rangle$ being the $x$-dependent ground state. Tuning $x$ from zero to one, we go from the SPT phase with the cluster GS to a trivial product state $|\psi_0(1)\rangle = |+\rangle^{\otimes N}$, through the critical point $x_{\mathrm{cr}} = 1/2$ where the GS develops nontrivial correlations. We highlight that one of the possible preparation circuits can be based on the Hamiltonian variational ansatz (HVA)  \cite{Wiersema2020PRXQ}. Due to the favorable dynamical Lie algebra scaling for this model \cite{WWHo2019SciPost} and large gradients for HVA in general \cite{park2023hamiltonian}, the corresponding feature map is efficient, meaning that the depth of the preparation circuit scales at most quadratically in the system size. However, for the purpose of numerical tests it is also instructive to use exact diagonalization, as it allows for clean studies of ground state embedding and its properties irrespective of variational preparation.
%%%
\begin{figure}[t!]
\includegraphics[width=1.0\linewidth]{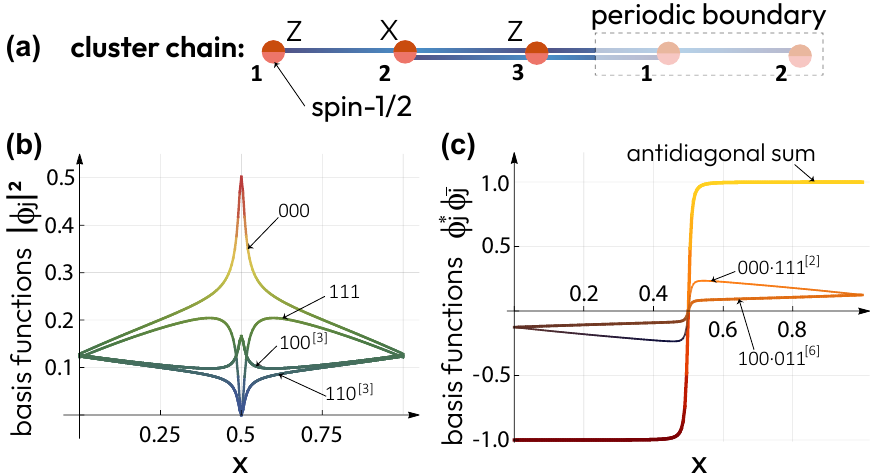}
    \caption{Analysis of a simple $N=3$ cluster model. \textbf{(a)} Sketch of the spin-1/2 chain with ZXZ couplings, periodic boundary conditions, and assuming the transverse magnetic field. \textbf{(b)} Basis functions of the ground state embedding for the $N=3$ cluster model, shown as squared projections $|\phi_j(x)|^2$ on computational basis states $j=000,100, ..., 111$. Superscript indices indicate degeneracies for one-hot and two-hot states. \textbf{(c)} Products of the basis functions $\phi_j^*(x) \phi_{\bar{j}}(x)$ corresponding to antidiagonal components, and their sum that represents the expectation value of the string order operator $\langle \hat{\mathcal{O}}\rangle(x)$.}
    \label{fig:cluster-basis}
\end{figure}
%%%

Finally, keeping in mind the Hamiltonian of interest, we stress that our goal is to distinguish the SPT phase from the trivial phase. For this we need to define the string order parameter $\hat{\mathcal{O}}$. Given that the 1D cluster model is based on $N$ 3-body stabilizers $\{ \hat{S}_i \}=\{ \hat{Z}_{i} \hat{X}_{i+1} \hat{Z}_{i+2} \}_{i=1}^{N}$, the string order corresponds to their product over the periodic boundary,
\begin{equation}
\label{eq:O-str}
    \hat{\mathcal{O}} = \prod_{i=1}^{N} \hat{S}_i = (-1)^N \hat{X}_1 \hat{X}_2 ... \hat{X}_N,
\end{equation}
and it corresponds to the parity operator with the reverted sign. Taking the expectation value of $\hat{\mathcal{O}}$ is our proxy to the definition of SPT order \cite{raussendorf2022measurementbased}, while other independent checks also include topological entanglement entropy estimation.
%%%
\begin{figure*}
\includegraphics[width=1.0\linewidth]{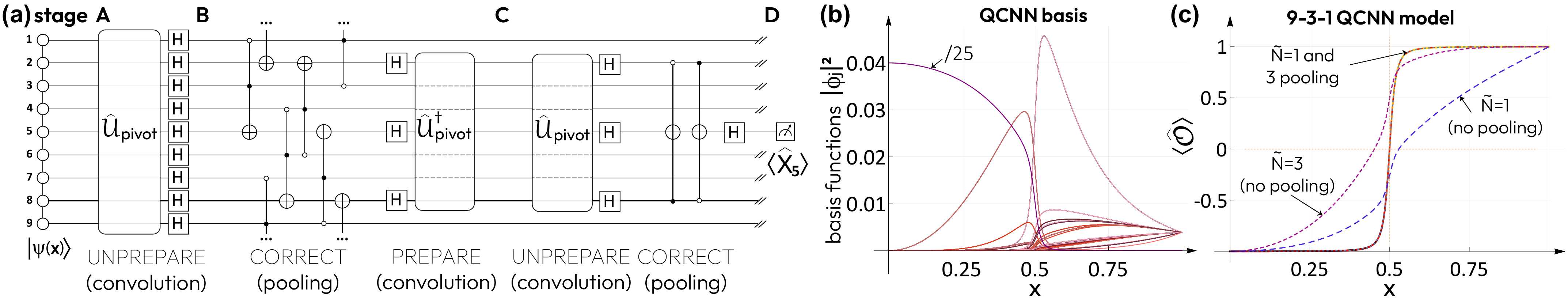}
    \caption{Analysis of the nine-qubit quantum convolutional neural network for the cluster state Hamiltonian. \textbf{(a)} QCNN circuit with a fixed predefined structure, motivated by the criteria for SPT phase recognition. It comprises of convolution and pooling layers, which represent unitaries that perform a basis change from the cluster state basis into a product basis (\textsf{UNPREPARE} operation corresponding to pivoting unitary $\hat{\mathcal{U}}_{\mathrm{pivot}}$ and a layer of Hadamards), and conditional operations that ensure that in the lowest order of symmetry breaking terms the product state remains trivial (\textsf{CORRECT} operation). We also label different stages of the protocol (stage A to D) that help understanding of how the underlying basis changes during QCNN processing. \textbf{(b)} Visualization of the basis set for $N=9$ QCNN at stage B, where squared amplitudes projected on computational basis states are shown as a function of the Hamiltonian parameter $x$. Multiple degenerate solution of the sigmoid type are generated. \textbf{(c)} The resulting models $\langle \hat{\mathcal{O}} \rangle (x)$ of the fixed QCNN that are read out at different stages. Orange curve corresponds to the full QCNN operation with pooling procedure, representing a sharp decision boundary at stage D for $\tilde{N} = 1$. This coincides with string order parameter measured full width 9-qubit $\langle \hat{X}_1 \hat{X}_2 ... \hat{X}_N \rangle$ prior to QCNN action, and $\tilde{N}=3$ measurement of the order parameter $\langle \hat{\mathcal{O}}_{\tilde{N}=3}\rangle \simeq \langle \hat{X}_2 \hat{X}_5 \hat{X}_8 \rangle$ at stage C. The case of measuring $\langle \hat{\mathcal{O}}_{\tilde{N}=1}\rangle \simeq \langle \hat{X}_5 \rangle$ and $\langle \hat{\mathcal{O}}_{\tilde{N}=3}\rangle$ observables in the absence of pooling (i.e. with \textsf{CORRECT} layers being removed) are shown with blue and magenta dashed curves. We note that here decision boundaries can be shifted by $\pm 1$ and/or trivially scaled by $1/2$ to yield the comparison where needed.}
    \label{fig:qcnn-931}
\end{figure*}
%%%

%-----

\subsection{Analysis of the ground state embedding from the QNN perspective} 

Once we have established the system Hamiltonian $\hat{\mathcal{H}}(x)$, the mapping procedure $\hat{\mathcal{U}}_{\varphi}(x)$, and the measurement (cost) operator $\hat{\mathcal{O}}$ [Eq.~\eqref{eq:O-str}], we proceed to classification from the point of view of quantum model building. For this, we study the basis set of our feature-dependent ground states $|\psi_0(x)\rangle$. Namely, we assume the embedded states to be decomposed in the computational basis $|\psi_0(x)\rangle \coloneqq \sum_j \phi_j(x)|j\rangle$, represented as computational $2^N$ basis states $\{ |j\rangle \}$ with corresponding binary/integer representation, weighted by the $x$-dependent basis functions $\{ \phi_j(x) \}_{j=0}^{2^N-1}$. It is instructive to visualize this basis in some form. We choose to project feature states onto the computational basis states (much like in case of QCBMs \cite{Coyle2020,kyriienko2022protocols}), reading out probabilities $|\phi_j(x)|^2 = |\langle j|\psi_0(x)\rangle|^2$, which can be seen as diagonal matrix elements of the corresponding density operator $\hat{\rho}_0(x) = 
|\psi_0(x)\rangle \langle \psi_0(x)|$. We choose a minimal example with $N=3$, shown in Fig.~\ref{fig:cluster-basis}(a), and plot the corresponding squared basis functions in Fig.~\ref{fig:cluster-basis}(b). We observe that all basis functions undergo a drastic change exactly at the critical point $x_{\mathrm{cr}}$, introducing an inductive bias for building models with an inherent criticality. We observe different behavior is associated to matrix elements that involve ferromagnetic bitstrings $j=000$ and $j=111$, one-hot bistrings ($j = 100, 010, 001$), and two-hot states $j=110,101,011$. Admittedly, these basis functions are shown in the computational basis, while the string order lies in the X Pauli plane. We next proceed to show how the required basis functions are ``picked up'' by the string order observable $\hat{\mathcal{O}} = (-1)^3 \hat{X}_1 \hat{X}_2 \hat{X}_3$. This specific string contains only anti-diagonal matrix elements (of $-1$ entries), such that the opposite (i.e. bitwise conjugated $j \leftrightarrow \bar{j}$) pairs of states are connected. In Fig.~\ref{fig:cluster-basis}(c) we plot the product of basis functions $\phi_j^*(x) \cdot \phi_{\bar{j}}(x)$ that are collected when estimating the expectation. One can see that all contributions represent sigmoid-like functions, with slightly increasing or decreasing fronts. Importantly, there are multiple degeneracies, such that one can pick multiple contributions when building the model (thus, success does not depend on specific projections and unique elements). Finally, we see that the sum of all antidiagonal products, being the expectation $\langle \psi_0(x) | \hat{\mathcal{O}} | \psi_0(x) \rangle$, represents a sharp decision boundary even at small system size. We proceed to see how this basis analysis plays out in the full QCNN workflow.

%------

\subsection{Analysis of QCNN basis transformation from the QNN perspective} 

Next, we want to see how the decision boundary is built by fixed quantum convolutional neural network with pre-defined convolution and pooling introduced in Ref.~\cite{Cong2019}. We remind that these are built based on the two criteria described previously in the text. To compose the fixed QCNN circuit we impose the criteria using the tools for lattice spin models. Specifically, one can observe that the pure ZXZ cluster Hamiltonian $\hat{\mathcal{H}}_{\mathrm{ZXZ}} = -J \sum\limits_{i=1}^{N} \hat{Z}_{i} \hat{X}_{i+1} \hat{Z}_{i+2}$ can be unitarily transformed into the trivial X Hamiltonian $\hat{\mathcal{H}}_{\mathrm{X}} = - h_{\mathrm{x}} \sum\limits_{i=1}^{N} \hat{X}_{i} $, with the transformation generated by the sum of nearest neighbour Ising terms $\hat{\mathcal{H}}_{\mathrm{ZZ}} = J_{\mathrm{zz}} \sum\limits_{i=1}^{N} \hat{Z}_{i} \hat{Z}_{i+1}$. This recently introduced procedure is referred to as \emph{pivoting} \cite{Verresen2023SciPost}. We observe that considering the pivot unitary
\begin{equation}
\label{eq:U_pivot}
    \hat{U}_{\mathrm{pivot}} = \exp\left( -i \frac{\pi}{4} \sum\limits_{j=1}^{N} (-1)^j \hat{Z}_{j} \hat{Z}_{j+1} \right),
\end{equation}
and assuming $J = h_{\mathrm{x}}$, we can perform the Hamiltonian transformation
\begin{equation}
    \hat{\mathcal{H}}_{\mathrm{X}} = \hat{U}_{\mathrm{pivot}} \hat{\mathcal{H}}_{\mathrm{ZXZ}} \hat{U}_{\mathrm{pivot}}^\dagger,
\end{equation}
and similar transformation applies to change the basis of corresponding eigenstates. The corresponding SPT ground state of ZXZ cluster model thus is unitarily connected with the trivial ground state of $|+\rangle^{\otimes N} \cong \hat{U}_{\mathrm{pivot}} |\psi_{0 \mathrm{(=SPT)}}\rangle$ (up to a global phase). Given the specific angle of $\pi/2$ two-qubit rotation, we can see pivoting as an effective layer of commuting CZ gates. Indeed, this is how cluster states are prepared in one-dimensional systems \cite{raussendorf2022measurementbased}.

We visualize the steps of the fixed QCNN in Fig.~\ref{fig:qcnn-931}(a) as the full circuit for $N=9$ qubits. The step-by-step description is presented in Appendix B, and here we summarize the main points. The fixed circuit structure represents the QCNN targeting the SPT phase recognition. For this, starting from the ideal cluster state at $x=0$ the goal is to de-entangle qubits via pivoting \cite{Verresen2023SciPost} (\textsf{UNPREPARE} layer in Fig.~\ref{fig:qcnn-931}a), and make sure that the expected value of the string order parameter is maximized. Additionally, assuming defects (bitflips $\hat{X}_i$) in the cluster state generated by the transverse field for $x>0$, the pooling layer is designed to test values of qubits that are traced out, and correct possible errors on qubits chosen for building the model (\textsf{CORRECT} layer in Fig.~\ref{fig:qcnn-931}a). 

Now, let us analyze the action of the fixed QCNN circuit in terms of building quantum models $f_{\theta_{\mathrm{fixed}}}(x)$ and track how the basis set of the function changes. We already know that initially (i.e. at stage A, Fig.~\ref{fig:qcnn-931}a) the basis corresponds to functions that peak and dip around $x = x_{\mathrm{cr}}$. Once we have performed the first pivot and arrived into the Z basis (stage B), we can plot the basis set, again representing as squared projections on computational basis states. This is shown in Fig.~\ref{fig:qcnn-931}(b). We see that the basis consists of functions which qualitatively match the switching-on and -off behavior, and contain many degenerate basis functions. %This helps to improve the performance, since while building the decision boundary we do not rely on picking very special combinations. 
Next, we proceed to the first pooling layer that corrects errors that may have arose on the subset of qubits (here qubits 2, 5, and 8). If we are to stop at the effective 3-qubit model, we shall return to the SPT ground state for this reduced register by applying \textsf{PREPARE} layer and measuring $\hat{\mathcal{O}}_{\tilde{N}=3} = (-1) \hat{X}_2 \hat{X}_5 \hat{X}_8$ at stage C in Fig.~\ref{fig:qcnn-931}a. We show the corresponding decision boundary as the $x$-dependent expectation $\langle \hat{\mathcal{O}} \rangle$ leading to the sharp transition in Fig.~\ref{fig:qcnn-931}(c) labeled as ``$\tilde{N}=1$ and $3$ pooling''. This overlays with the decision boundary based on the full $N=9$-qubit string order parameter measured prior to the action of QCNN (blue dashed curve). Next, we continue to another QCNN pooling layer, such that the model is reduced from 9 to 1 qubit. In this case we unprepare the 3-qubit SPT ground state, and perform checks on qubits 2 and 8 such that the state of qubit 5 matches the expected value. Essentially, it corresponds to measuring $\hat{\mathcal{O}}_{\tilde{N}=1} = -\hat{X}_5$ (stage D in Fig.~\ref{fig:qcnn-931}a). We plot the corresponding decision boundary and see that it again matches the sharp transition in Fig.~\ref{fig:qcnn-931}(c) labeled as ``$\tilde{N}=1$ and $3$ pooling''. However, what if there are no pooling layers applied to the input data state, and we measure reduced size operators directly after the convolutional layers? The models for $\tilde{N}=3$ and $\tilde{N}=1$ are shown in Fig.~\ref{fig:qcnn-931}(c) as dashed violet and magenta curves, labeled with ``(no pooling)'' tag. We see that in this case the decision boundary is blurred, and cannot offer the same degree of accuracy, mostly due to deviations in the critical region.

Finally, let us offer another understanding of the fixed QCNN workflow. The action of 9-3-1 QCNN leads to the model being $\langle \psi(x)| \hat{\mathcal{U}}_{\mathrm{QCNN}}^\dagger \hat{X}_5 \hat{\mathcal{U}}_{\mathrm{QCNN}} |\psi(x)\rangle$, where $\hat{\mathcal{U}}_{\mathrm{QCNN}}$ collect all layers (notice that no mid-circuit measurements are required, and we can simply trace out anything but the middle qubit). We can then see the action of QCNN as measuring an effective ``dressed'' operator $\hat{\mathcal{U}}_{\mathrm{QCNN}}^\dagger \hat{X}_5 \hat{\mathcal{U}}_{\mathrm{QCNN}} \cong \hat{X}_1 \hat{X}_2 ... \hat{X}_9$, which is equal to the string order parameter up to the global phase. This simply means that pooling makes sure we pick up the relevant basis states from our embedding, while building the model on the small subset of qubits. This is very important for trainability, as we reduce the shot noise (smaller number of measurements is required for models based on single-qubit sampling), and still enjoy the access to the full basis. We believe that the very same strategy can be extended beyond the QCNN and applied for building other QML models, and QCNN-like measurement adaptation can improve solvers beyond classification.  

%-------------

\section{Training QCNNs}

Departing from the fixed QCNN structure, we now allow for adjustable elements and compose a variational quantum convolutional neural network keeping the same overall QCNN layout. 
%%%
\begin{figure}[t!]
\includegraphics[width=1.0\linewidth]{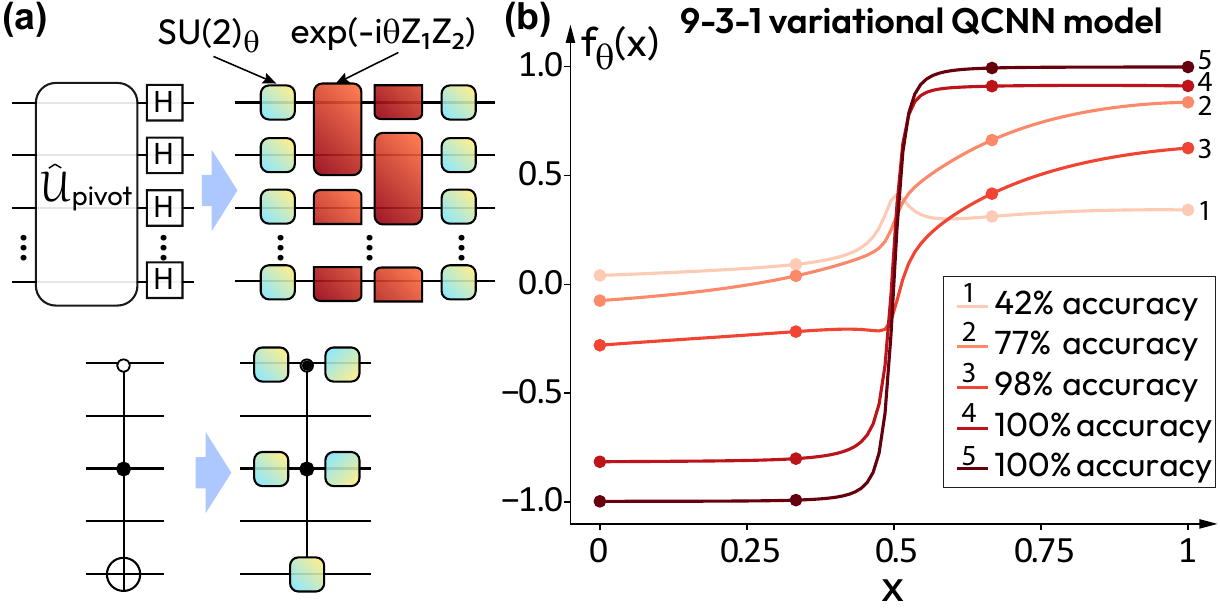}
    \caption{\textbf{(a)} Parameterized elements of the variational QCNN circuit that substitute the fixed QCNN elements for pivoting and correction layers. Layers are translationally invariant but SU(2) and ZZ unitaries have different angles. \textbf{(b)} Examples of the decision boundary for the 9-qubit variational QCNN during the training. We show decision boundaries, starting from random initialization (curve 1; 42\% test accuracy), and increasing number of epochs to 25, 50, 100, and 200 (curves 2-5, respectively).}
    \label{fig:var-QCNN}
\end{figure}
%%%

%---------

\subsection{QCNN trained on cluster ground state embedding}\label{Variational_QCNN}

We proceed to compose the variational QCNN circuit. The goal is to build a variational ansatz inspired by the fixed QCNN structure to see whether we can recover a suitable basis for classification via training. We start by dealing with the convolutional layer. First, we change the layer of Hadamard gates into a layer of trainable SU$(2)_\theta$ gates, where we implement an arbitrary one-qubit rotation as a sequence of three fixed-axis rotations, SU$(2)_\theta = \hat{R}_{\mathrm{X}}(\theta_1)\hat{R}_{\mathrm{Z}}(\theta_2)\hat{R}_{\mathrm{X}}(\theta_3)$. We also make the pivot unitary adjustable. This corresponds to changing the pivot unitary based on ZZ$(\pi/2)$ gates to arbitrary angle two-qubit operations ZZ$(\theta)$. We decide to place an SU(2) layer before and after the pivot layer to allow for more control over the basis set. The same three parameters are used for each of the SU$(2)_\theta$ gates in a layer to ensure translational invariance, and the same applies for each ZZ$(\theta)$ gate, as Fig.~\ref{fig:var-QCNN}(a) shows. Hence, the convolutional layer acting on the full nine-qubit register has seven independent parameters. We also note that the convolutional layers acting on three qubits depicted in Fig.~\ref{fig:qcnn-931}(a) can be cancelled out, and the final Hadamard is swapped out for an SU$(2)_\theta$ gate. 
We then turn to the pooling layer, where we again use controlled operations to avoid the use of mid-circuit measurements and show that QCNN effectively corresponds to an efficient basis adaptor. The fixed Toffoli gates are changed into trainable gates, where the target is now acted on by the SU$(2)_\theta$ unitary. To enable bit flipping of the control gates, we place one-qubit SU$(2)_\theta$ gates before each control qubit, and the corresponding SU$(2)_\theta^\dagger$ gate after each control qubit. We again keep translational invariance by ensuring each trainable Toffoli gate within the pooling layer has the same parameters. Therefore, the pooling layer has nine independent parameters. 

The input state is the $x$-dependent GS of the cluster Hamiltonian $|\psi(x)\rangle$. The QCNN circuit is applied on this state, and a final measurement $\hat{O} = \langle\hat{X}_5\rangle$ determines the output of the model. Hence our model $f_\theta(x)$ can be written in the form given by Eq.~\eqref{eq:QCNN-model}. Our goal is to perform the binary classification of the two phases corresponding to the symmetry-protected topological order (class A) and the staggered ferromagnetic order (class B). To train this model, we sample training data points $\{x_\alpha\}$ uniformly between $0$ and $1$, with labels determined by measuring the SPT order parameter, and assigning $y_\alpha$ labels $-1$ and $1$ to class A and class B, respectively. 
%$\hat{O}$ on the input states:  $y_\alpha = \langle \psi(x_\alpha) | \hat{\mathcal{O}} | \psi(x_\alpha) \rangle$. 
Our loss function is the MSE as defined in Eq.~\eqref{eq:QCNN-MSE-loss}, and we minimize this loss using stochastic gradient descent. Specifically, we use the Adam optimizer with a learning rate of $0.05$. We then test the trained model on data $\{x_\beta\}$ randomly sampled from a normal distribution. This is to ensure proper testing of the trained QCNN around the critical point, where classification is most challenging. 
When testing the binary classification, we evaluate the model for each $x_\beta$. If $f_\theta(x_\beta) < 0$, then we place $|x_\beta\rangle$ in class A, corresponding to the SPT phase, and if $f_\theta(x_\beta) > 0$, then $|x_\beta\rangle$ is in class B, corresponding to the trivial phase. Hence even if $f_\theta(x_\beta)$ is not close to the actual test label $y_\beta$, as long as the sign of the model output equals the sign of the test label, the data point $x_\beta$ is considered correctly classified. Test accuracy is the measure of the proportion of test data points that are placed into the correct class. 
%%%
\begin{figure}[t]
\includegraphics[width=1.0\linewidth]{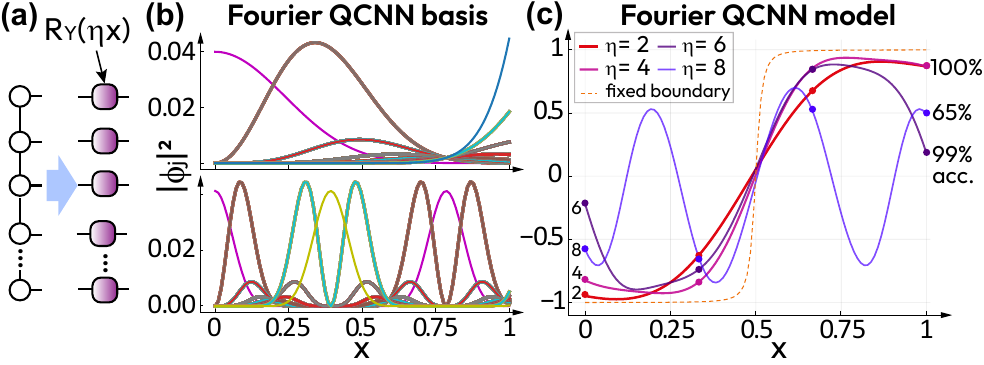}
    \caption{\textbf{(a)} Here, we substitute the ground state embedding by the rotation-based embedding with Fourier basis. $\eta$ represents a parameter that defines the frequency range. \textbf{(b)} Basis sets shown for the Fourier feature map with $\eta = 2$ (top panel) and $\eta = 8$ (bottom panel), visualized as squared projections on the computational basis, $|\phi_j(x)|^2$. The state is taken after the embedding is performed, just before the variational QCNN circuit. \textbf{(c)} Decision boundary coming from QCNN with the Fourier feature map based on $\hat{R}_{\mathrm{Y}}(\eta x)$ rotations, shown for $\eta=2,4,6,8$. The corresponding test accuracy for each embedding is shown on the right, ranging from $65\%$ ($\eta = 8$) to $100\%$ ($\eta = 2$ and $4$).}
    \label{fig:Fourier-QCNN}
\end{figure}
%%%

This is demonstrated in Fig.~\ref{fig:var-QCNN}(b), where we trained a model on 4 training points (shown as dots) and tested it on 100 test data points, from which we plotted the curves $f_\theta(x)$. Curve 1 depicts the QCNN output at initialization, i.e. with random untrained parameters. Even at this initial stage, we can observe a small jump around $x=0.5$, hinting at some change of behaviour at the critical point. Hence even with non-optimized QCNN parameters, the GS feature map contains sufficient information to indicate certain critical behavior. Curves 2 and 3 in Fig.~\ref{fig:var-QCNN}(b) are plotted for the QCNN model trained for 25 and 50 epochs, respectively. We can see that these decision boundaries do not match what we require (the fixed QCNN boundary seen in Fig. ~\ref{fig:qcnn-931}c), but the shape of the curves and the behaviour at criticality is becoming more accurate with increased training. So much so, that the test accuracy for curve 3 in Fig.~\ref{fig:var-QCNN}(b) is almost at 100\%; the values of $f_\theta(x_\beta)$ are incorrect, but most of the test points still fall on the correct side of 0, which is sufficient for the binary classification task. Next, curve 4 (Fig.~\ref{fig:var-QCNN}b) is plotted for the trained QCNN model after 100 epochs. At this stage 100\% accuracy has been achieved, but the boundary still does not coincide with the fixed QCNN boundary. Finally, after 200 training epochs we obtain curve 5. With enough training we do recover the correct, sharp decision boundary. Therefore, the GS feature map with a suitably chosen variational QCNN performs very well for state classification, generalizing to unseen data even with very few training data points. 

It is important to note that even with an arbitrarily chosen variational ansatz, we can still get excellent results for classification using the GS feature map, as we demonstrate in Appendix \ref{apx_arbitrary_ansatz}. This reinforces that it is the specific choice of feature map that provides the right basis for finding a sharp decision boundary.

%---

\subsection{QCNN trained on rotation-based embedding}

We have seen that we can train a QCNN model to perform effectively the classification of ground states for the chosen Hamiltonians. But what happens if we go from the GS feature map to a more generic embedding protocol? We demonstrate this by using rotation-based embedding which leads to the conventional Fourier-type quantum model. We apply a layer of single qubit Pauli rotation gates acting on the zero state as the new feature map (Fig.~\ref{fig:Fourier-QCNN}a). The resulting input state reads $|\psi(x)\rangle = \prod_{i=1}^N \hat{R}^i_{\mathrm{Y}}(\eta x)|{\o}\rangle$, where $\eta$ is a fixed parameter, and again $N=9$. Changing $\eta$ changes the frequency range that the model has access to, and this manifests itself in the basis sets depicted in Fig.~\ref{fig:Fourier-QCNN}(b). As we did for the GS feature map to produce Fig.~\ref{fig:qcnn-931}(b), we prepare $|\psi(x)\rangle$ before projecting onto the computational basis states to measure the $x$-dependent diagonal components of the input density operator, $|\phi_j(x)|^2$. In contrast to the GS basis, we do not observe any rapid change in behaviour at the critical point; rather, this Fourier embedding induces a sinusoidal behaviour in each of the basis functions. There is a significant difference observed as you increase $\eta$. At $\eta = 2$ there are fewer ``dominant'' functions (top panel in Fig.~\ref{fig:Fourier-QCNN}b), meaning that low frequency models can be constructed. However, at $\eta = 8$ the basis looks more oscillatoric, with many peaks appearing due to the high frequency of the feature map  (bottom panel in Fig.~\ref{fig:Fourier-QCNN}b).

These feature-map-induced basis changes have a clear impact on the QCNN classification, as Fig.~\ref{fig:Fourier-QCNN}(c) shows. We train the QCNN circuit as we did previously, with the same training and test data set, loss and optimizer choices. We can see that for a low-frequency Fourier map ($\eta = 2,4$), the QCNN is roughly able to find the fixed QCNN boundary. The transition is less sharp at the critical point, and $f_\theta(x)$ does not quite reach -1/+1 for $x$ close to 0/1, but the decision boundary gives a clear separation between the two phases, and hence ensures 100\% accuracy on the classification task. However, as $\eta$ increases further, the impact of the high-frequency basis set becomes clearer, as it becomes harder to pick the basis functions for fitting the correct boundary. This culminates in the result for $\eta = 8$; the many ``dominant'' sinusoidal peaks in the basis cause the QCNN boundary to also resemble a sinusoidal curve. Naturally, this has an impact on test accuracy. In this case many test points are situated in the upper left and bottom right quadrants of the plot, indicating incorrect classification.

What does the study of QCNN models with Fourier embedding tell us? One take-home message is the importance of setting the frequencies. We can also see it as the necessity of feature pre-processing, or in the context of quantum kernels adjusting the kernel bandwidth \cite{Shaydulin2022kernels}. In cases which allow feature engineering, QCNNs with rotation-based embeddings may still form relatively high-performing models. However, for tasks that do not have this option the generalization can be poor. %(e.g. being sensitive to the introduced bias in analysis)

We also stress that our analysis considers the `test accuracy' as a success measure for classification typically adopted in quantum machine learning \cite{Li2021rev}. However, there are various measures beyond the test accuracy that can provide a stronger separation between models based on different embeddings. For instance, this includes precision recall scores, confusion matrix, receiver operating characteristic curves, and many others. In this sense, our result show that even for largely forgiving metrics the separation is significant, and is only expected to grow for other parameters and training regimes.

%-----

\subsection{Generalization}\label{generalization_sec}

Next, we study the generalization property for classification for different embeddings. Specifically, we show that QCNNs with the ground state embedding can generalize well (have small generalization error \cite{Caro2022}) as compared to the rotation-based embedding.

\begin{figure}[t!]
\includegraphics[width=1.0\linewidth]{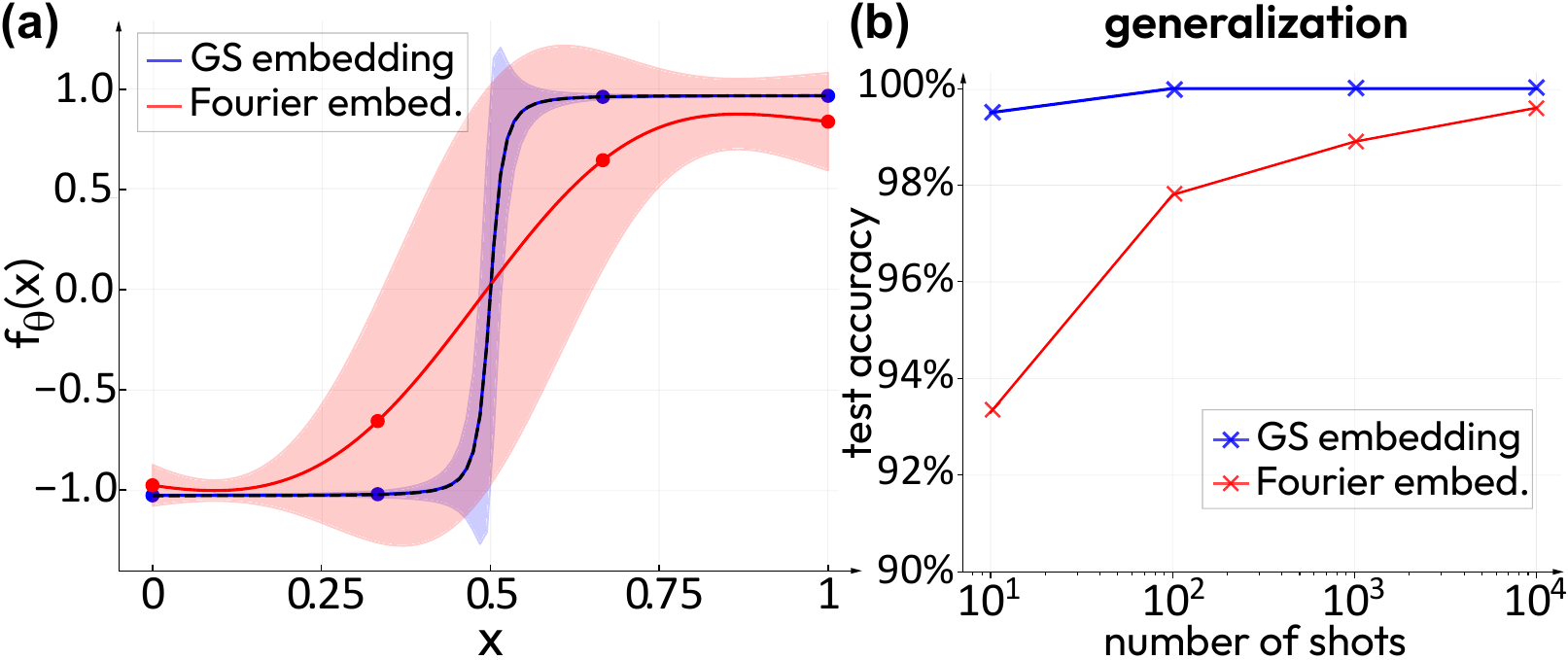}
    \caption{\textbf{(a)} Decision boundaries for the two types of embeddings (ground state and Fourier rotation-based), for a finite number of shots. Shaded area shows the corresponding variance, where large variance leads to less accurate results. \textbf{(b)} Generalization measure as a function of the number of measurement shots.}
    \label{fig:generalization}
\end{figure}

The results so far assumed access to an infinite number of shots when measuring expectation values, enabling us to evaluate $f_\theta(x)$ exactly. The separation in the performance between the GS and Fourier embeddings is even more apparent when we move to the finite shot regime. The error $\epsilon$ in the estimation of $\langle\hat{O}\rangle$ is given by $\epsilon = \sqrt{\mathrm{var}(\hat{O})/N_s}$, where $\mathrm{var}(\hat{O}) = \langle\hat{O^2}\rangle-\langle\hat{O}\rangle^2$ and $N_s$ is the number of shots. Fig.~\ref{fig:generalization}(a) depicts the decision boundaries for the GS feature map and the Fourier feature map ($\eta = 2$), with the blue and red curves representing $f_\theta(x)$ in the infinite shot limit (ground state and Fourier embeddings, respectively). The shaded blue and red area shows the region [$f_\theta(x)$-$\mathrm{var}(f_\theta(x))$, $f_\theta(x)$+$\mathrm{var}(f_\theta(x))$] for both types of embedding.
%%%
%%%

We can see that for both embeddings, the variance away from the critical point is relatively low compared to the variance around the critical point. In these low-variance regions, an accurate measurement of $\langle\hat{O}\rangle$ can be found even with very few shots. Towards the critical point the variance becomes increasingly large, and many more measurement shots are required. We note that the shaded regions for the QCNN with Fourier embedding are much larger than those of the QCNN with GS embedding. The Fourier model has a region of error that spans both sides of the $f_\theta(x)=0$ threshold for a much wider range of $x$. 

This manifests itself in the results seen in Fig.~\ref{fig:generalization}(b). For this, we consider a minimal scenario for showing the separation between the two QCNN model types. Specifically, we train a model assuming access to an infinite number of shots, but when calculating the test accuracy we measure $f_\theta(x)$ by averaging over a fixed number of measurements. We repeat this procedure 10 times for both the GS and Fourier embeddings, and we plot the average test accuracy across the 10 trials as a function of number of shots. We see that test accuracies still remain high for both feature maps, but a real separation is observed for low numbers of measurement shots. Even for points away from the critical point, the high variance of the Fourier QCNN model means that 10 shots is not enough to evaluate $f_\theta(x)$ with any real degree of accuracy, making the model susceptible to incorrect classifications. Finally, 10,000 shots are required to return to the average test accuracy across the 10 trials in the infinite shot regime (99.6\%). Meanwhile, the GS QCNN model only struggles with points very close to the critical point in the low shot regime. In fact, 100 shots is already enough to return to the 100\% test accuracy found across all 10 trials with an infinite number of shots.

These results further demonstrate the benefit of the GS feature map. The sharp transition at criticality ensures that errors arising from access to a finite number of shots have a minimal impact on classification.  In Appendix \ref{trained_with_noise}, we also show that the GS embedding maintains this advantage in learning a sharp decision boundary when considering a model both trained and tested with shot noise. The smooth decision boundary is transformed into a noisy transition line, and the lower sharpness of the Fourier QCNN is further exposed in this setting. Even more importantly, considering cases where data is limited (e.g. when only few runs of quantum hardware can be analyzed due to time constraints), the generalization on few training examples and read out with few samples can largely yield the classification for actual devices.

%==========

\subsection{Solving regression tasks with QCNNs}

Given what we have learnt from classification with quantum convolutional neural networks, we suggest to investigate the performance of QCNNs for specific regression tasks. In particular, we test it for problems containing sharp transitions, using QCNN using both ground state feature map and Fourier feature map.

To solve a regression problem we consider a target function $f(x)$ that describes a relationship we want to find in a dataset. A trial function or ``surrogate model'' $f_\theta(x)$ can express a family of functions dependent on what the variational parameters $\theta$ are. The model can be trained by minimizing the MSE loss function
\begin{align}
    \mathcal{L}(\theta) = \frac{1}{M}\sum_{\alpha=1}^{M} \Big\{f_\theta(x_\alpha) - f(x_\alpha)\Big\}^2,
\end{align}
where $\{x_\alpha\}_{\alpha=1}^M$ is the set of training points and $M$ is the cardinality of this set. Similarly to the previous section we build the quantum model $f_\theta(x)$ based on QCNN expectation value, and additionally include classical variational parameters to rescale and shift the expectation.
%%%
\begin{figure}[ht]
\includegraphics[width=1.0\linewidth]{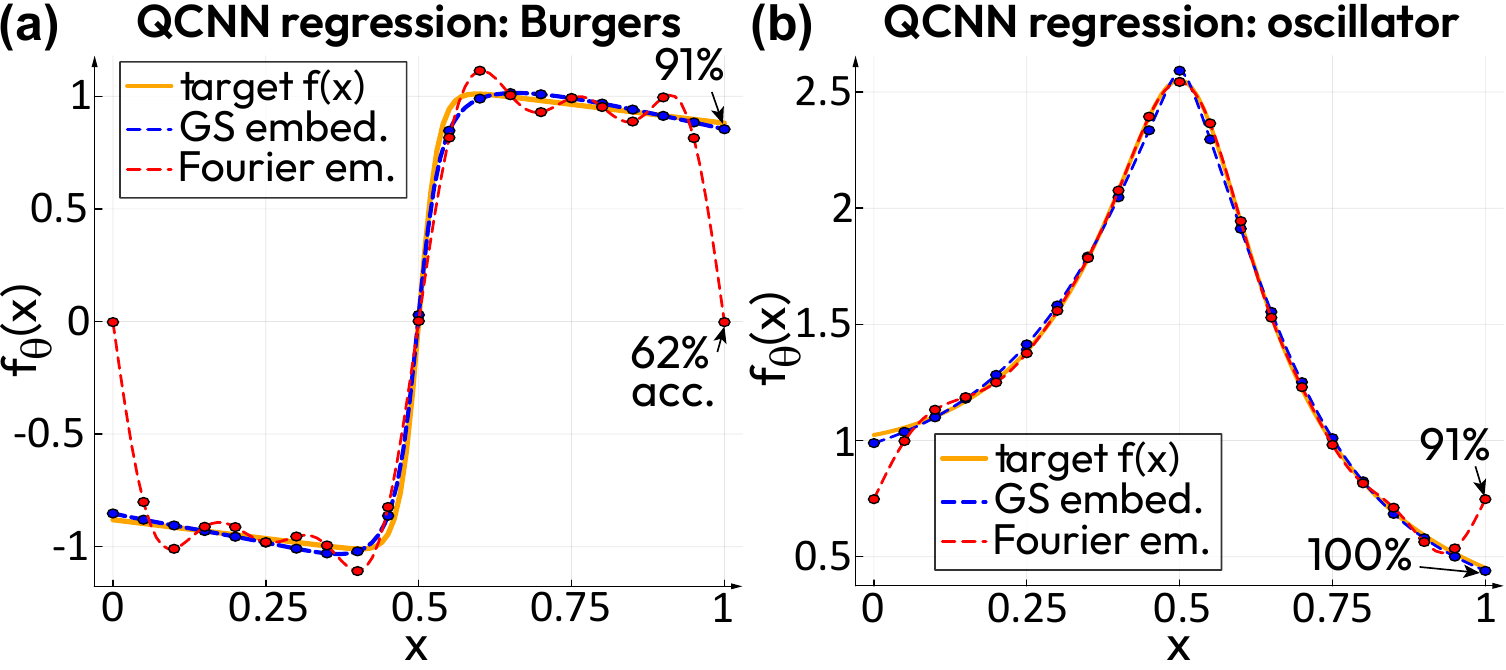}
    \caption{\textbf{(a)} Viscous flow solution of the Burgers equation learnt by the quantum convolutional neural network with the ground state embedding and rotation-based Fourier embedding. Test accuracy, calculated as percentage of points within square error less than $0.05$, is estimated to be 91\% for the GS model and 62\% for the Fourier model. \textbf{(b)} Transfer function of a single degree of freedom oscillator learnt by QCNNs with the two feature maps, together with corresponding accuracies.}
    \label{fig:regression}
\end{figure}
%%%

First, we consider an example coming from fluid dynamics. As a target function for training we choose a solution of Burgers equation, assuming the regime that features a critical behavior. The corresponding partial differential equation reads \cite{anderson1995computational}
\begin{align}
    \frac{\partial u}{\partial t} + u \frac{\partial u}{\partial x} =  \nu \frac{\partial^2 u}{\partial x^2},
\end{align}
with the boundary conditions
\begin{align}
    u(0,x) &= -2 \frac{\nu}{\phi(0,x)}\frac{d\phi}{d x} + 4, ~~~ x \in [0, 2 \pi], \\
    u(t, 0) &= u(t, 2 \pi), ~~~ t \in [0, T], \\
    \phi(t, x) &= \exp\left\{-\frac{(x-4t)^2}{4 \nu (t+1)}\right\} + \exp\left\{-\frac{(x-4t-2\pi)^2}{4 \nu (t+1)}\right\},
\end{align}
the solution can be found as $u(t,x) = -2 [\nu/\phi(t,x)] d\phi/dx + 4$.
%
%\begin{align}
%    u(t,x) = -2 \frac{\nu}{\phi(t,x)}\frac{d\phi}{d x} + 4.
%\end{align}
%
We rescale this solution and consider the initial time $t=0$, setting the target function as $f(x) = [u(0, \pi + 1/2 - x) - 4]/3$ with $\nu = 0.05$. 
We remind that our goal is to learn the known solution of the differential equation, thus performing regression by QNN training, also known as quantum circuit learning \cite{mitarai2018quantum}. Additionally, we note that one can potentially adopt the workflow to match a physics-informed approach with feature map differentiation \cite{kyriienko2021solving}, and we leave the question of differentiating hidden feature maps for future works.

We perform regression selecting $M=21$ equally-spaced training points over $[0,1]$, for both the ground state embedding based on cluster Hamiltonian (same as in the previous section) and the Fourier feature map based on rotations $\hat{R}^i_{\mathrm{Y}}(2\pi x)$ for each qubit $i$. Training is performed via Adam optimizer, and we return to the state vector simulation to compare the performance of the feature maps in the absence of shot noise. The results are as shown in Fig.~\ref{fig:regression}(a). We observe that the original QCNN with GS embeddings is able to represent the transition with high degree of generalization, and the Fourier model experiences oscillations while roughly representing the trend. We assign an accuracy metric for each model, corresponding to the percentage of test points for which the square error difference between the trained function and target function is with an error bound $\epsilon$. The test points are chosen as 100 uniformly spaced points and error bound as $\epsilon = 0.05$. The accuracy of the QCNN model with cluster Hamiltonian embedding is then equal to 91\%, mostly due to slightly different scaling of tails, and the Fourier-based QCNN has only 62\% test accuracy due to poor generalization. We conclude that ground state based embedding with the known region of criticality and critical exponents can be translated to problem in other domains of physics, offering distinct model-building capabilities with good generalization. 

The second example that we consider corresponds to a transfer function of a single degree of freedom oscillator with damping \cite{stafford2017scenario}. Specifically, the transfer function reads
\begin{align}
    H(x) = \frac{1}{\sqrt{\left(1-x^2 \right)^2 + \left(2\zeta x \right)^2}},
\end{align}
where $x$ denotes the frequency of the oscillator normalized by its natural frequency, and $\zeta$ is the damping ratio. We rescale the transfer function as $f(x) = H(1.6[x + 0.1])$ to match the range of QCNN-based models, and set $\zeta = 0.05$ such that the transition behaviour occurs centred around $0.5$. 
Regression is performed the same way as the viscous Burgers case, as well as the test accuracy estimation. The results are as shown in Fig.~\ref{fig:regression}(b). We observe that both the ground state feature map and Fourier feature map are able to approximately fit the shape, with the GS embedding outperforming Fourier due to better generalization at the end point regions. %Beyond regression on data, we can also use the described QCNN models for solving differential equations using the physics-informed quantum models \cite{kyriienko2021solving}.

Overall, from the use of QCNNs for regression we can see that it is important for the basis functions from the choice of encoding to match the problem. This is particularly important for embedding choices based on ground states, where all the basis functions have a strong particular behaviour at the critical point. This application of the QCNN with GS embedding to regression tasks is a form of transfer learning; the specific basis functions induced by the feature map allows the quantum model to perform particularly well for fitting a particular class of functions. Additionally, one can make the GS feature map become variational as well, such that the critical region and sharpness are altered as needed. Once a suitable embedding choice is set, the QCNN can be used to represent the target function efficiently and learnt with proven trainability \cite{pesah2021absence} --- an important advantage over fully-connected quantum neural networks.

Finally, we highlight that there are many physical problems based on quantum measurements that go beyond classification and assume inference of some system properties, thus favoring regression. For instance, this can be estimation of magnetization and specific heat for magnetic materials, charge order in superconductors etc. In this case we suggest that QCNNs can play an important role due to their naturally-emerging basis set with inductive bias and high generalization.

%====================

\section{Discussion}

There are several crucial aspects to be recalled and discussed. First, we stress that QCNNs are indeed very successful in solving problems for cases where input states are generated with the basis set that suits classification. Here, we can say that the advantage in both machine learning and quantum machine learning is underpinned by data --- while functions that represent decision boundaries are not necessarily difficult to fit, the generalization comes from embeddings that are natural to considered problems (e.g. classification of physical system behaviour). Thus, we suggest that similar advantage in terms of generalization can be attained when solving regression tasks for physics-informed problems.

We made the observation that QCNNs working with quantum data are typically built on feature maps with an inductive bias \cite{bowles2023contextuality} that favors the problem. Given that QCNN circuits have a number of independent adjustable parameters at the post-processing stage, which is not prohibitively large, they are practically trainable and in conjunction with geometric QML \cite{Larocca2022PRXQ,JJMeyer2023PRXQ,nguyen2022theory} approaches they can provide superior performance in terms of both classification, anomaly detection \cite{kyriienko2022unsupervised}, and regression.

Second aspect concerns addressing problems with multiple features. This assumes embedding multivariate functions as decision boundaries. To date, mostly few-variable cases were considered for physical systems with the ground state embedding \cite{Caro2022}, and the properties of QML models based on these ground state feature maps have been recently studied \cite{umeano2024ground}. Meanwhile rotation-based feature maps were routinely used for datasets with a large number of features \cite{Li2021rev}. Once we want to use ground state embeddings for multiple features, we face a task of assigning each feature $x_\ell$ to some Hamiltonian parameter. One option is to use different type of Hamiltonian terms (interaction range, coupling etc) to encode features $x_\ell$, but this implies that critical behavior for different features is distinct. Another option is composing Hamiltonians with inhomogeneous couplings, for instance with 1D cluster models (each described by $x_\ell$) collated into a ladder. How can we use QCNNs to process data with many features? This question has to be answered for extending the use of QCNN beyond few physical examples. 

Third aspect concerns the type of measurement adaptation circuit. If we consider pooling being a way to adopt the measurement basis from few-body to effectively $N$-body operators, can we do it with only a limited gate set? And if we follow this route, is it sufficient to use matchgate circuits, which offer both control and classical simulability \cite{kasture2022protocols}? Or can we do the adaptation with Clifford-only circuits \cite{Mitarai2022PRRes}? If this is the case, the advantage of QCNN has to come directly from data, assuming that embedding of features corresponds to non-trivial quantum processes that cannot be probed otherwise. For such processes one can consider a highly beneficial scenario where pre-training of QCNNs is performed \textit{in silico} classically for limited number of states. Then, we can run QCNNs as measurement adaptation circuits for physical devices with imperfections, performing classification with small number of samples and enjoying the corresponding advantage.

Another important question regarding the power of QCNNs and possible embeddings arises when we consider extensions from simple one-dimensional models and go into two dimensions. Namely, every feature can be in principle embedded into ground states of 2D spin models with distinct topological properties. In this case, the entanglement structure, correlations, and nature of phase transition change. And so does a critical exponent of the transition \cite{zeng2018quantum}. For instance, considering the toric code model or models with fractional statistics, one can enrich the basis for describing critical phenomena \cite{Wen2017RMP}. Can we match them with corresponding behaviour in hydrodynamical systems, thus ensuring quantum ``simulation'' of phase transitions in fluid dynamics? Can we extend it to chaotic systems? The development of exotic embeddings with long-range topological order may well offer capabilities in classification and regression beyond currently explored states with limited entanglement.

Finally, QCNNs have been shown to have a strong connection to error correction, where pooling layers can be used to perform syndrome extraction and denoising \cite{Cong2019}. At the same time, we suggested to look at the process via explicit modelling paradigm. Can we extend it to some of the known error correction protocols, explaining them as ``sharpening'' of models and feature engineering? These are the questions that can be addressed adopting the proposed strategy.

We must also acknowledge the recent works by Cerezo \textit{et al.} \cite{cerezo2023provable} and Bermejo \textit{et al.} \cite{bermejo2024qcnn}. These works detail how some of trainable parameterized quantum circuits, in particular QCNNs, generally live in a polynomial subspace of operators. This opens up the possibility of their efficient classical treatment in the Heisenberg picture \cite{goh2023liealgebraic}. At the same time, working with non-trivial input states and specifically quantum data represents a beneficial scenario. %This draws into question whether this architecture can lead to a tangible quantum advantage in any setting. 
Additionally we note that these studies suggest a possibility for QCNNs either being initialized or wandering out of this polynomial subspace during training. %Feasibly there could be problems (involving non-trivial quantum datasets) where one needs to enter the exponentially large subspace in order to optimize the model. 
In this situation, full quantum simulation of the QCNN is necessary. Maintaining the trainability guarantees is also important, hence the entanglement of these quantum states must be limited. Finding such datasets is an open challenge.

%========================

\section{Conclusion}

In this work we have developed a better understanding of working with quantum data. Dissecting a workflow of quantum convolutional neural networks as an example, we identify reasons for their success in tasks involving quantum data, particularly quantum phase recognition. We have shown that supplied quantum states (features) can be understood in terms of \emph{hidden} feature maps --- quantum processes that prepare states depending on classical parameters $\bm{x}$. During the mapping process, we observed that the ground state preparation supplies a very beneficial basis set, which allows the building of a nonlinear quantum model for classification with decision boundaries that are sharp and generalize from only few data points and measurement samples. We show that single-qubit observables in QCNNs can be used to ``pick up'' the most beneficial basis functions. This can lead to advantage in using a small number of samples, as compared to measure-first approaches with shadow tomography. The developed understanding also opens another perspective on quantum sensing aided by convolution-based quantum processing. Finally, motivated by classification, we applied the QCNN-based workflow with cluster model ground state embedding to solve problems in fluid dynamics (viscous Burgers equation) and wave physics (damped oscillator). This suggests quantum convolutional neural networks as a provably trainable tool for data-driven modeling of critical phenomena with quantum computers.

\begin{acknowledgements}
We thank Lachlan Lindoy and Ivan Rungger for useful discussion on the subject. O.K. and C.U. and acknowledge the funding from UK EPSRC (award EP/Y005090/1), as well as support from NATO SPS project MYP.G5860.
\end{acknowledgements}

\appendix

\section{Ground state preparation}
    
Here, we briefly summarize different strategies to prepare QCNN inputs as ground states or low-temperature ensembles of states. First, we note one can prepare 
\begin{equation}
    |\psi_0(x)\rangle = \mathcal{T} \left\{ \exp\Big[-i \int ds \hat{\mathcal{H}}(s; x)\Big] \right\}
\end{equation}
using an appropriate adiabatic preparation path for Hamiltonian $\hat{\mathcal{H}}(x)$ (here $\mathcal{T}$ denotes time-ordering) \cite{AlbashLidar2018RMP}. This however depends on the properties of the Hamiltonian and gap closing. The adiabatic approach also allows quantum models based on ground-state feature maps to be analysed as Fourier models \cite{umeano2024ground}.

Second, one can consider the effective thermal state preparation process
\begin{equation}
    \hat{\rho}_{\mathrm{th}}[\hat{\mathcal{H}}(x)] = \lim\limits_{\beta \rightarrow \infty} \frac{\exp(-2\beta \hat{\mathcal{H}}(x)}{\mathrm{tr}\{\exp(-2\beta \hat{\mathcal{H}}(x))\}},
\end{equation}
where $\beta$ is an effective inverse temperature. This can be simulated approximately with quantum imaginary time evolution (QITE) techniques \cite{Motta2020}.

Third, given that cluster model can be overparametrized in linear (or at most quadratic) depth, we can assign a ground state preparation unitary with pre-optimized angles $\Phi_{\mathrm{GS}}$ such that there is a unitary $\hat{\mathcal{U}}_{\Phi_{\mathrm{GS}}}[\hat{\mathcal{H}}(x)]$ that prepares $|\psi_0(x)\rangle$ up to sufficient pre-specified precision. We suggest to use the QAOA-type preparation \cite{WWHo2019SciPost}, which corresponds to the Hamiltonian Variational Ansatz (HVA) \cite{Wiersema2020PRXQ} for the considered cluster Hamiltonian. Let us consider $J_{\mathrm{xx}}$ for brevity, and use the transverse field cluster model as an example. The feature map for the ground state embedding can be composed as
\begin{equation}
\label{eq:psi-HVA}
    |\psi_0(x)\rangle = \prod\limits_{d=1}^{D} \left( e^{-i \Phi_{d_1}^{\mathrm{(opt)}}(x) \sum_{i \in \mathcal{S}} \hat{Z}_i \hat{X}_{i+1} \hat{Z}_{i+2}}  e^{-i \Phi_{d_2}^{\mathrm{(opt)}}(x) \sum_{i} \hat{X}_i} \right) |{\o}\rangle,
\end{equation}
assuming that we apply sufficient number of layers $D$, and optimal angles $\{ \Phi_{d_{1,2}}^{\mathrm{(opt)}} (x) \}$ are recovered for all system features $x$. We note that this is possible for $D$ that enables overparametrization \cite{larocca2023theory}, and was shown to give exact ground state preparation for integrable models \cite{WWHo2019SciPost,bespalova2021quantum}.

The ground state preparation strategy will determine the quantum resources (circuit depth, ancilla qubits) required to implement the ground state feature map, as well as the quality of the quantum model; the smooth basis functions visualised in Fig.~\ref{fig:qcnn-931}(b) become noisy around the critical point \cite{umeano2024ground}.

%----

\section{Fixed QCNN analysis}

We visualize the steps of fixed QCNN in Fig.~\ref{fig:qcnn-931}(a) of the main text as the full circuit for $N=9$ qubits. Let us go step by step aiming to understand the underlying machine learning operations. Starting with the prepared ground state for the pure cluster model ($x=0$ point), we observe that pivoting brings the system to $|+\rangle^{\otimes N}$ and the layers of Hadamard gates make it the computational zero state $|{\o}\rangle$. We call this as \textsf{UNPREPARE} operation layer. Thus at $x=0$ we satisfied the second QCNN criterion (discussed in Section II.B of the main text), meaning that at the pooling stage (Fig.~\ref{fig:qcnn-931}a, stage B) we would get 0 bit measurements deterministically on $2N/3$ qubits, while keeping $\tilde{N} = N/3$ qubits (we choose the pool of qubits $2,5,8$). The criterion 1 (i.e. fixed point one) can be satisfied by re-preparing the cluster state now on the three selected qubits by applying Hadamard and $\hat{U}_{\mathrm{pivot}}^\dagger$ layer. We get the smaller version of $|\psi_0\rangle_{N/3}$. Next we need to consider the case of $x>0$, where the deviation from the zero magnetic field generates ``errors'' as bitflips. Namely, operations $\{ \hat{X}_i \}$ can be applied to on any qubit line on top of ideal symmetry protected topological (SPT) ground state $|\psi_0(0)\rangle$, and propagate through the circuit. That is what we want to correct when performing the pooling procedure, trying to keep the clean copy of SPT phase as far as possible. We observe that $\hat{U}_{\mathrm{pivot}} \hat{X}_i |\psi_0\rangle = \hat{Z}_{i-1} \hat{X}_i \hat{Z}_{i+1} \hat{U}_{\mathrm{pivot}} |\psi_0\rangle$ and $\hat{\mathrm{H}}^{\otimes N} \hat{U}_{\mathrm{pivot}} \hat{X}_i |\psi_0\rangle = \hat{X}_{i-1} \hat{Z}_i \hat{X}_{i+1} \hat{\mathrm{H}}^{\otimes N} \hat{U}_{\mathrm{pivot}} |\psi_0\rangle$, and these are the effective errors that we need to correct. Given that in the absence of errors pivoting and X-to-Z basis map generates $|{\o}\rangle$, we required to correct states of the type $|0...01010...0\rangle$ (and cyclic shifts), and make sure that after pooling $|0\rangle^{\otimes N/3}$ state is recovered. To avoid mid-circuit measurements we suggest to use the recipe of deferred measurements and assign controlled operations that are compatible with measuring (and tracing out/discarding) qubits that are not pooled through. This can be achieved by applying Toffoli gates $\hat{\mathrm{CCX}}_{i,j,k} \coloneqq \mathbbm{1}_{i,j,k} - |1\rangle
_i\langle 1|\otimes |1\rangle
_j\langle 1| \otimes \mathbbm{1}_{k} + |1\rangle
_i\langle 1|\otimes |1\rangle
_j\langle 1| \otimes \hat{X}_{k}$, where $i$ and $j$ are control qubit indices, and $k$ represents a target. We also define the bit flipped version of this gate $\hat{\mathrm{CCX}}_{i,j,k}^{(01)} \coloneqq \hat{X}_i \hat{\mathrm{CCX}}_{i,j,k} \hat{X}_i$, with the $i$-th control now corresponding to 0 state. This gate is denoted by open circle when conditioned on 0 and not 1. Our idea is detecting the patterns of bits in next-nearest neighbors, such the when distance-one flipped pair is detected, several layers of Toffolis correct the qubits 2, 5, and 8 (chosen as targets; see Fig.~\ref{fig:qcnn-931}a, \textsf{CORRECT} layers). One can the check that errors propagating to measurement do not impact the observable, in the lowest order.

%----

\section{QCNN with arbitrary variational ansatz}\label{apx_arbitrary_ansatz}
In our previous studies we used a variational ansatz inspired by the fixed QCNN structure; this architecture proved to be extremely successful in recovering an accurate decision boundary for the cluster Hamiltonian. One question may arise; can we obtain similar success if the variational ansatz is chosen more arbitrarily, without guidance from the fixed (non-variational) QCNN structure. To investigate this, we select a more general variational ansatz, taken from \cite{Hur_2022}. Specifically, we select arbitrary SU(4) gates acting on pairs of neighbouring qubits to make up the convolutional layers, and controlled rotations as the pooling layers (still maintaining the 9-3-1 structure and translational invariance). 

\begin{figure}
    \centering
    \includegraphics[width=\linewidth]{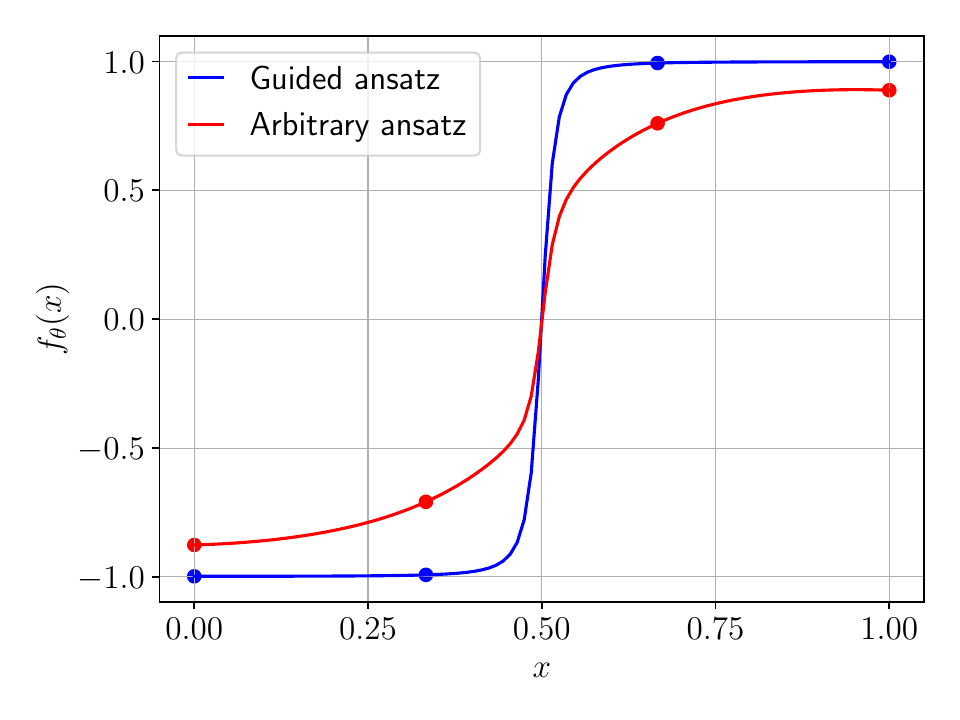}
    \caption{Decision boundaries for the QCNN with GS embedding with the original, "guided" ansatz detailed in Sec.~\ref{Variational_QCNN}, and an arbitrary variational ansatz.}
    \label{fig:arbitrary_ansatz}
\end{figure}

We train the model in the same way as before, and we compare the performance of the GS embedding QCNN with the guided and arbitrary variational ansatze in Fig.~\ref{fig:arbitrary_ansatz}. We find that $100\%$ test accuracy is still achieved with this arbitrary ansatz trained on 4 training points, and the sharpness at criticality is maintained. Again, the importance of the specific basis of the GS embedding is on show.
%%%
%%%

\section{QCNN trained with shot noise}\label{trained_with_noise}
In Sec.~\ref{generalization_sec}, we demonstrated that sampling from the QCNN with GS embedding would generally lead to more accurate results, as the high variance of the Fourier embedding model induces errors in the finite shot regime. Here, we extend the analysis by both training and testing the model with a fixed number of shots. We apply the same procedure as before, just this time we train (and test) the model with 200 shots per expectation value evaluation.
%%%
\begin{figure}
    \centering
    \includegraphics[width=\linewidth]{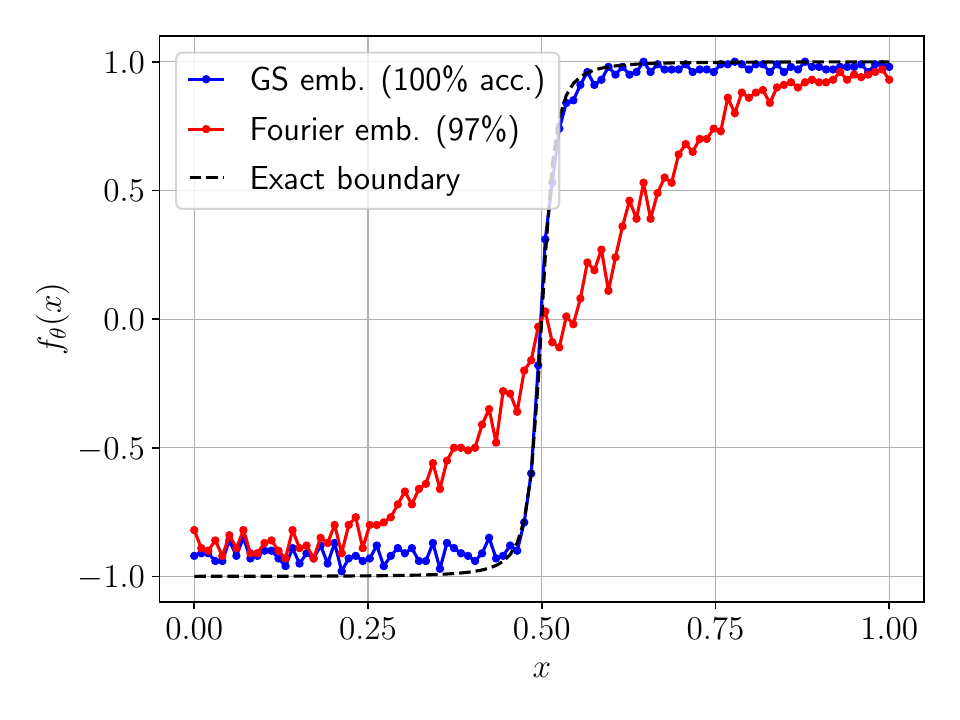}
    \caption{Decision boundaries for the QCNN with GS and Fourier embeddings, trained and tested with 200 shots per expectation value estimate. Dots represent the test data points: the errors in the critical regime for the Fourier embedding are apparent.}
    \label{fig:train_finite_shots}
\end{figure}
%%%

As Fig.~\ref{fig:train_finite_shots} shows, the QCNN model with GS embedding is still more effective in this regime. Although the tails of the GS feature map boundary are particularly noisy and do not quite match the exact boundary, the smooth, sharp transition at criticality is still maintained in the presence of shot noise. Meanwhile, the decision boundary for the QCNN with Fourier embedding is noisy throughout. Most importantly, the shot noise induces classification errors for data points around the critical region; only 97\% test accuracy, compared to 100\% for the GS embedding. 

%\clearpage

%\newpage

%\bibliography{bibliography}

\input{main_v2.bbl}

\end{document}